\documentclass{llncs}
\usepackage{graphicx}
\usepackage{booktabs}
\usepackage{authblk}
\usepackage{import}
\usepackage{appendix}
\usepackage{tabularx}
\usepackage{hyperref}
\usepackage{url}

\usepackage{pgfplots}
\pgfplotsset{compat=1.8}
\usepgfplotslibrary{statistics}
\begin{document}

\title{On legitimate mining of cryptocurrency in the browser -- a feasibility study}

\author{Saulius Venskutonis\inst{1} \and
Feng Hao\inst{2} \and
Matthew Collison\inst{1}
}

\institute{Newcastle University, Newcastle upon Tyne, NE1 7RU, UK
\\
\email{\{s.venskutonis1, matthew.collison\}@newcastle.ac.uk},
\and
University of Warwick, Coventry CV4 7AL, UK
\\
\email{Feng.Hao@warwick.ac.uk}
}
\date{29 November 2018}

\maketitle              

\noindent
\makebox[\linewidth]{\small 29 November 2018}

\begin{abstract}
Cryptocurrency mining in the browser has the potential to provide a new pay-as-you-go monetisation mechanism for consuming digital media over the Web. However, browser mining has recently received strong criticism due to illegitimate use of mining scripts in several popular websites (a practice called “cryptojacking”). Here we provide the first feasibility study of browser mining as a legitimate means of monetisation in terms of revenue, user consent and user experience within a specially built website. Our results compare browser mining to display advertisement and indicate browser mining provides a preferable user experience to advertising when the hash rate is user-adjustable. Furthermore, over 60\% of participants would select browser mining over advertisement if they were invested in the ecosystem by obtaining half of the mined cryptocurrency. Our estimations show that browser mining currently generates revenue at a rate 46 times less than advertisement, however we would expect that gap to decrease as we observed a significant drop in mining difficulty after our tested cryptocurrency implemented ASIC-resistant mining measures. Overall, based on our results we find browser mining to be a legitimate alternative to display advertisement and conclude by discussing its current limitations and potential applications.
\end{abstract}


\section{Introduction}
The first widely used library to mine cryptocurrency in the browser arguably emerged on the 14th of September, 2017 with the release of Coinhive's JavaScript miner \cite{coinhive}. Since then browser-based mining has been gaining publicity and media attention \cite{coinhive_google_trends}, mostly due to unethical attempts to profit from unsuspecting website visitors in hacked popular media \cite{cbs_showtime_hack,politifakt_hack} and more recently, government websites \cite{india_hack}. In a short span of time unethical use of browser mining in the wild has harmed its reputation, resulting in many major anti-virus providers regarding it as a malicious activity in need of restriction \cite{malwarebytes_blocks_mining,kaspersky_blocks_mining}. However, due to the accessibility and privacy-protective nature of browser mining \cite{first_look_into_cryptojacking}, it provides a mechanism to confidentially monetise practically any digital media stream or web page, which many web services may find appealing. For example, ``The Pirate Bay'' have previously experimented with browser mining in an effort to replace advertisements \cite{pirate_bay_mining}.

Currently, display advertisement remains the principal method of monetisation of online content \cite{ads_expenditure,us_ads_expenditure}. 
However, whilst the cost of advertisement has risen by an average of 12\% over the last two years, advertisers find it significantly less valuable than expected \cite{adobe_digital_insights}. Furthermore, about 11\% of Web users are using ad-blocking software and its use is predicted to grow further \cite{adblock_statistics}, whilst subscription pricing models are increasingly being favoured by high profile digital content providers such as Spotify and Netflix. This raises the question of whether the dominance of advertisement could potentially be challenged by alternative means of monetisation with a more flexible pay-as-you-go access to digital media and websites.

Ethical practice is key to legitimate monetisation and browser mining could not be considered a legitimate alternative to advertisement without established mechanisms of collecting informed user consent \cite{first_look_into_cryptojacking}. In an attempt to bring legitimacy to browser mining Coinhive developed the ``AuthedMine'' JavaScript miner, which requires explicit user consent to operate. This certainly indicates progress towards legitimisation of browser mining but leaves dense perplexity: whilst a growing number of internet users reject digital ads via ad-blockers, it is unclear if they would instead consent to browser mining or if they would fully comprehend what they are consenting to \cite{first_look_into_cryptojacking}.

Therefore, in light of novel methods for cryptocurrency mining becoming available to the public we conducted the first user study, which evaluated the feasibility and instinctive user choice between ads and browser mining on both desktop and mobile clients and surveyed the participants to gain further insight into their decision rationale. In this paper, we present the results of this study to discuss limitations and the future of browser mining.


\section{Methodology}
To conduct an anonymised user study, we built an experimental online blog, https://www.hippocrypto.me, which utilised Coinhive to mine privacy-focused cryptocurrency "Monero" \cite{monero_wp}. The website was distributed online to form a convenience sample of 107 volunteers between ages 18-50 (71.03\% male, 27.1\% female, and 1.87\% other). We restricted access to the website, which allowed participants to view its content only after advertisement or browser mining was explicitly selected as a monetisation method (participants could make their choice only once). To obtain consent, participants were presented with an opt-in screen, which largely resembled the message currently used by Coinhive \cite{coinhive_authedmine}. Afterwards, they were provided with a survey, which asked for demographic information (age, gender, level of education), familiarity with cryptocurrency, and their view towards browser mining. Participants were also asked to indicate the reasons for their monetisation choice and, assuming they selected advertisement, whether they would be willing to select browser mining next time if they could keep 50\% of the mined cryptocurrency.

In addition, we collected website usage data during multiple sessions to evaluate behavioural trends. Session lengths were measured as the time between accessing the content and leaving the website. To assure data anonymity each individual received a RFC4122 \cite{id_rfc4122} compliant unique ID stored in local storage on each browser, which was used to identify participants and aggregate data from multiple sessions. We also parsed browser User-Agent strings to estimate whether a mobile or desktop device was used. Furthermore, Coinhive's miner can be adjusted to run at varying degrees of power referred to as \textit{throttle}. To evaluate user behaviour at different levels of CPU usage, at the start of each session we applied a degree of A/B testing and set the miner to randomly initiate at either 10\%, 50\% or 80\% throttle (with 10\% being the minimal throttle allowed). Participants were able to continuously monitor the active throttle value via an integrated dashboard in the side menu and were instructed how to adjust it at any point.

We estimated revenues from both monetisation methods using variables listed in Table \ref{table: ads_revenue_vars}. For estimation purposes the following assumptions were made. Firstly, to estimate the advertisement revenue, this study assumed a fixed price for two differently-sized banner advertisements, which would be variously priced in real-world conditions \cite{ads_pricing}. In addition, advertisement networks generally utilise several different pricing metrics \cite{ads_pricing_models}, however in this study only the cost-per-thousand impressions (CPM) was used to estimate ad revenue. In practice, users would be able to interact with advertisement potentially adding to the revenue via clicks or actions. CPM is generally dynamic, therefore the value used in our calculations (\$2.80) is an estimated average derived from multiple reports \cite{cpm_average} referred here as \(CPM_{avg}\). Considering Google AdSense is currently the dominant advertising network \cite{adsense_share_img}; when estimating advertisement revenue \(R_{ads}\) we used its fee \cite{adsense_fee} to calculate the net CPM income for the publisher \(P_{rpm}\). The total number of advertisement impressions was computed by keeping a counter and incrementing it by the number of adverts on each page whenever on-site navigation occurred. For browser mining, Coinhive's API was used to obtain the total number of hashes submitted on each session, which was then divided by the session length to infer the average hash rate. The conversion rate, mining difficulty and block reward were recorded on 17th April 2018. Finally, the collected data was forwarded to an external analytics server where it could be aggregated and visualised.



\section{Results}
\subsection{Revenue}

Our estimates show browser mining generated revenue at a rate 46x less than advertisement (\(R_{adsMax}\) compared with \(R_{cryptoMax}\) in table \ref{table: r_cryptoMax=r_adsMax}). Furthermore, over the period of the experiment the total revenue from the participants who chose mining (23 of 107 individuals) was nearly 75x less than the revenue produced by the majority that chose advertising. Assuming the number of users and the average session time remained constant, the average hash rate per user would have to increase from 63 h/s to approximately 2738 h/s for revenue to be comparable. To put in perspective, currently there are no common consumer devices capable of achieving such hash rates. On the other hand, if the hash rate was to remain the same the session length would have to increase from over two minutes to nearly two hours for mining revenue to be equivalent.


\begin{table}[h!]
\caption{Values used to estimate revenues, including Monero values on 17th of April, 2018 \cite{monero_april_17}. }
\begin{tabularx}{\textwidth}{lXl}
\toprule
Variable & Description & Value \\
\midrule
I & Total ad impressions during the experiment & 118 \\
$CPM_{avg}$ & Revenue per 1,000 impressions \cite{cpm_average} & \$2.80 \\
$X_{a}$ & AdSense fee & 32\% \\
$P_{rpm}$ & Net revenue per 1,000 impressions & \$1.904 \\
$M_{usd}$ & Value of Monero in USD & \$193.44 \\
D & Monero mining difficulty & 57377400931 \\
B & Block reward & 4.87 \\
$X_{c}$ & Coinhive fee  & 30\% \\
H & Total hashes submitted during the experiment & 260,864 \\
M & Total Monero mined during the experiment & 0.0000154 \\
\bottomrule
\end{tabularx}%
\label{table: ads_revenue_vars}

\end{table}



\begin{table}[h!]
\caption{Estimated ad impression, hash rate, session length, and revenue values. Maximum revenue is an estimation assuming all participants were to select the same monetisation method. }
\resizebox{\textwidth}{!}{%
\begin{tabular}{lll}
\toprule
Variable & Description & Value \\
\midrule
$A_{avg}$ & Average ad impressions per user & 2.512 \\
$T_{avg}$ & Average session length when mining & 151.732 s \\
$U_{avg}$ & Average hash rate & 63.675 h/s \\
$R_{ads}$ & Total revenue from advertisement & \$0.224 \\
$R_{crypto}$ & Total revenue from cryptocurrency mining & \$0.00299 \\
$R_{adsMax}$ & Estimated maximum ads revenue during experiment & \$0.511 \\
$R_{cryptoMax}$ & Estimated maximum mining revenue during experiment & \$0.011 \\
U & Avg. hash rate required for mining revenue to match ads & 2738.593 h/s \\
T & Avg. session length required for mining revenue to match ads & 6525.829 s \\
\bottomrule
\end{tabular}%
}
\label{table: r_cryptoMax=r_adsMax}

\end{table}

\subsection{User consent}
The collected data suggests people were more inclined to select advertisement, with only 21.5\% (n=23) of participants selecting cryptocurrency mining as their preferred monetisation option. When asked why advertisement was selected the two most prominent reasons indicated were concerns about reduced device performance (25.84\%, n=46) and a higher level of familiarity with advertisement (23.6\%, n=42). In contrast the leading reason for selecting browser mining was a dislike for advertisement (36.59\%, n=15). Only 1.69\% (n=3) of all participants who selected advertisement, did so due to a negative view of cryptocurrency. It is worth noting that questions which inquired for decision rationale, were multiple choice, meaning a single participant was able to indicate more than one reason for selecting a particular monetisation method. Segregating responses into groups of mobile and desktop or laptop users revealed only a small portion of participants on mobile devices have chosen cryptocurrency mining (14.81\%, n=4). Reasons for selecting advertisement within this subgroup included concerns about reduced battery life (26.79\%, n=15), diminished device performance (21.43\%, n=12) and better familiarity with advertisement (19.64\%, n=11).

The observed census showed a general inclination towards advertisement across all demographic factors, with female participants notably comprising only 6.9\% (n=2) of the miners. Further investigation revealed lack of familiarity with browser mining as the leading reason (46.51\%, n=20) for elevated female choice of advertisement and concerns about device performance coming second (23.26\%, n=10).

Advertisement was the preferred choice across all levels of self-indicated familiarity with cryptocurrency. The highest rate was seen amongst participants who indicated a slight familiarity, 88.89\% (32 out of 36) of whom selected advertisement, while the highest rate of browser mining was among highly familiar participants (12 out of 26, 46.15\%). Furthermore, we observed that as the levels of familiarity increased so did the number of browser-mining instances, except for participants who held the highest level of claimed understanding (experts), none of whom selected browser-mining. Analysing cryptocurrency familiarity with attitude towards browser mining revealed experts held only neutral, negative, or strongly negative opinions. In contrast, the remaining familiarity groups ranging from the lowest (no familiarity) to the second highest (high familiarity), were mainly neutral and expressed at least some level of positivity towards browser mining. 60.71\% (n=51) of participants, who selected advertisement indicated willingness to substitute their monetisation choice if half of the mined cryptocurrency was reimbursed as an incentive, 27.45\% (n=14) of whom held a
negative and 3.92\% (n=2) a strongly negative view towards browser-based cryptocurrency mining.


\subsection{User experience}
Throughout the study session length was recorded and analysed under the assumption that users would terminate sessions earlier if user experience was affected by the monetisation method. The median session length was found to be 6x higher when browser mining compared to advertisement. Splitting results by device revealed a similar trend; when mining the median session lengths on desktop and mobile devices were 7.8x and 2x longer respectively (Fig.~\ref{fig: avg_sess_box_plot}). Longer median session lengths with browser mining were observed across all levels of familiarity with cryptocurrency. High familiarity resulted in the longest sessions with mining, whilst moderate familiarity produced lengthiest median sessions with advertisement.

Analysis of data segregated by gender indicated that with browser mining female participants had 2.3x shorter median sessions than their male counterparts whilst with advertisement the median session lengths remained relatively similar across all genders. Additionally, a 7x increase in median session length when mining compared to advertisement was observed among participants with bachelor's degree as the highest level of achieved education. Participants with master's or doctoral degrees did not follow a similar trend, however their sample size was significantly smaller (73.83\% of participants had bachelor's degrees). Similarly, the 18-24 age range showed a 5.9x higher median session length on browser mining versus advertisement.

Evaluation of median session times regarding the initial browser mining throttle indicated that the longest sessions were produced by the lowest starting throttle (10\%), with the highest throttle (80\%) coming second. We also measured the average change in the initial throttle value (fig.~\ref{fig:avg_throttle_change}), which revealed that generally the increase was more significant than the reduction, however participants were willing to increase the throttle value only if was initially small.

\begin{figure}[!htp]
    \begin{minipage}[t]{.5\linewidth}
    \begin{tikzpicture}[baseline={(0,0)}]
      \begin{axis}
        [
        legend entries = {Mobile/Ads (M/A), Mobile/Mining (M/M), Desktop/Ads (D/A), Desktop/Mining (D/M)},
        legend style={
        legend pos=north west,
        font=\tiny
        },
        width=1.10\textwidth,
        height=5.5cm,
        xtick=\empty,
        ytick=data,
        boxplot/draw direction=y
        ]
        \addlegendimage{no markers,black}
        \addlegendimage{no markers,red}
        \addlegendimage{no markers,green}
        \addlegendimage{no markers,orange}
        \addplot+[
        black,
        boxplot prepared={
          median=8.005,
          upper quartile=54.521,
          lower quartile=4.493,
          upper whisker=54.521,
          lower whisker=2.376,
          sample size=23,
        },
        ]
        coordinates {}
        node[above,inner sep=1pt,font=\tiny] at
(boxplot box cs: \boxplotvalue{median},0.5)
{\pgfmathprintnumber{\boxplotvalue{median}}}
;
        \addplot+[
        red,
        boxplot prepared={
          median=16.025,
          upper quartile=45.858,
          lower quartile=6.863,
           upper whisker=45.858,
          lower whisker=4.578,
          sample size=4,
        },
        ]
        coordinates {}
        node[above,inner sep=1pt,font=\tiny] at
(boxplot box cs: \boxplotvalue{median},0.5)
{\pgfmathprintnumber{\boxplotvalue{median}}}
;
        \addplot+[
        green,
        boxplot prepared={
          median=7.4,
          upper quartile=31.091,
          lower quartile=3.867,
           upper whisker=31.091,
          lower whisker=0.949,
          sample size=61,
        },
        ]
        coordinates {}
        node[above,inner sep=1pt,font=\tiny] at
(boxplot box cs: \boxplotvalue{median},0.5)
{\pgfmathprintnumber{\boxplotvalue{median}}}
;
        \addplot+[
        orange,
        boxplot prepared={
          median=58.324,
          upper quartile=144.653,
          lower quartile=23.614,
           upper whisker=144.653,
          lower whisker=2.152,
          sample size=19,
        }
        ]
        coordinates {}
        node[above,inner sep=1pt,font=\tiny] at
(boxplot box cs: \boxplotvalue{median},0.5)
{\pgfmathprintnumber{\boxplotvalue{median}}}
;
      \end{axis}
    \end{tikzpicture}
    
    \caption{Session time in seconds. Upper whiskers were omitted to preserve formatting. Their values are: M/A=330, M/M=49.3, D/A=17483, D/M=1299.4}
\label{fig: avg_sess_box_plot}
\end{minipage}
    \begin{minipage}[t]{.5\linewidth}
    \begin{tikzpicture}[baseline={(0,0)}]
    \begin{axis}[
    width=1.10\textwidth,
    height=5.5cm,
    ybar,
    stack negative=separate,
    xtick=data,
    xticklabel={\pgfmathprintnumber\tick\%},
    enlargelimits=0.2,
    nodes near coords={\pgfmathprintnumber\pgfplotspointmeta\%},
    nodes near coords align={vertical}
    ]
    \addplot coordinates {(10,20) (50,-5) (80,-2.556)};
    \end{axis}
    \end{tikzpicture}%
    \caption{Average change in the three starting mining throttles}
    \label{fig:avg_throttle_change}
\end{minipage}
\end{figure}

\section{Discussion}
The results of this study concur with the recent research by Papadopoulos et al. \cite{truth_about_cryptomining} in that there is a significant gap between revenues from advertisement and browser mining, which depends on the device's computation power and the amount of time spent mining. Based on our estimates we assume matching advertisement revenue by solely increasing the hash rate is practically impossible with current consumer devices. Additionally, study data depicts limiting factors, which we think are currently imposed on browser mining. Firstly, lack of exposure to legitimate browser mining practices have naturally put most participants in a neutral position and swayed them towards what they perceived as a risk-averse choice. Recent discreditation of browser mining could deter from further adoption thus limiting exposure to legitimate browser mining practices and maintaining the status quo. Secondly, considering the increasing usage of smartphones, which can be depicted by the fact mobile is expected to overtake desktop devices in advertisement expenditure \cite{ads_expenditure}; clear avoidance of browser mining amongst mobile users could be discouraging to publishers. Furthermore, since scalability of browser mining is significantly restricted as noted by Papadopoulos et al. the rivalry for computational power would be fiercer thus reducing potential profits for competing publishers. In addition, attracting large reputable publishers may require an independent governing body to set standards ensuring safe and fair use.

However, if the observed increase in the consumption of online content has a causal relationship with browser mining and is not simply a result of experienced novelty, browser mining could potentially have some interesting use-cases. For example, it could constitute a pay-as-you-go monetisation model for web applications or media streaming services. Our results indicate that the revenue from browser mining could still be lesser than from advertisement assuming the expected average session length \cite{avg_session_video_streaming,how_long_ppl_stay_on_page}, however unobstructed access to immediate monetisation along with enhanced privacy could be sufficient benefits for some publishers to embrace browser mining regardless. Perhaps if the miner throttle remained within adequate bounds (which in this study were between 30\% and 77\%) and the number of concurrent browser mining instances was limited to one, a suitable balance between user experience and revenue opportunity could be established.

\begin{figure}
    \centering
    \includegraphics[width=\textwidth]{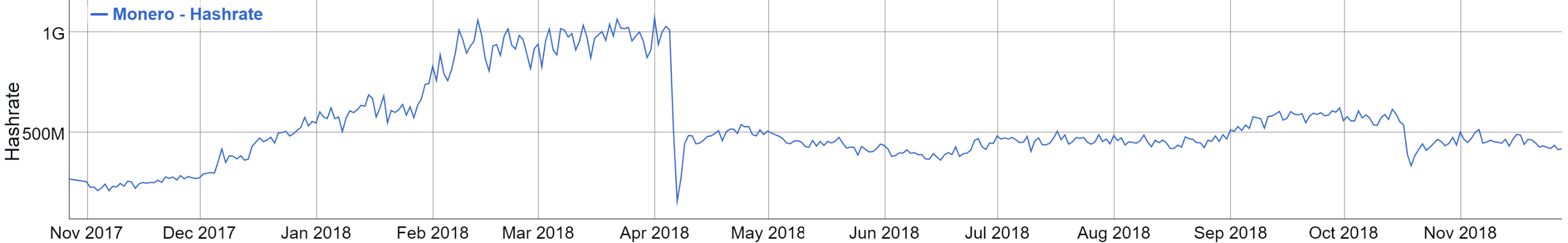}
    \caption{Monero global network hash rate significantly plummeted on 7th of April, 2018}
    \label{fig:monero_hashrate_change}
\end{figure}

It is also noteworthy to mention that during our experiment one of the largest producers of application-specific integrated circuit (ASIC) miners "Bitmain", announced its release of the first ASIC-powered miner \cite{bitmain_asic_monero}, designed to mine the CryptoNight hashing algorithm used by Monero. ASIC miners can achieve drastically higher hash rates than conventional CPU and GPU devices, since they are specialised to mine specific hashing algorithms. Use of ASIC miners in other cryptocurrencies, such as Bitcoin has been openly criticised \cite{asic_critique_bitcoin} due to their potential of centralising the mining power in the hands of selected few large mining pools. CryptoNight algorithm has been long thought to be "ASIC-resistant" by requiring extensive amounts of fast memory to be available on the hardware, thus making ASIC devices economically infeasible to produce \cite{why_monero_is_asic_resistant}. To combat Bitmain's potential ASIC threat, on April 6th, 2018 Monero's development team performed an emergency update to its protocol (practice known as a "hard fork"), which resulted in an almost 80\% loss of hash power and mining difficulty within its network (fig. \ref{fig:monero_hashrate_change}). This sparked theories that ASIC mining has been present in Monero's network even before the release of Bitmain's ASIC miner \cite{monero_has_asic_miners}. Thus, hypothetically with the significant reduction in mining difficulty and the potential exclusion of ASIC competitors; browser mining could have become more economically viable after the fork.

\noindent
Whilst browser mining may not be a complete replacement to advertisement, recent research along with our results indicate potential opportunities for it to compete with advertisement in niche markets. Considering digital advertisement's annual expenditure is \$215.8 billion \cite{ads_expenditure} and the fact that browser mining could potentially enter this highly valuable market, more research should be conducted to evaluate its performance on monetising various types of content using large sample pools.



\section{Conclusion}
In this paper we evaluated whether browser mining could be a legitimate monetisation alternative to advertisement by conducting the first study, which measured behavioural features, including user consent to mine cryptocurrencies in the browser and the session lengths resulting from both monetisation methods. Our estimations demonstrate that at the moment advertisement remains unchallenged in terms of revenue for the publisher and prospects of mobile use, however based on our results we suggest that browser mining can be a legitimate monetisation mechanism and its current notorious reputation could be overstated. Whilst there are significant ethical and legal avenues left to explore, we believe that given its wide range of possible legitimate use cases this interesting new concept deserves further research attention.

\bibliographystyle{splncs04}
\bibliography{references}


\end{document}